\documentclass[lenghtcheck,twocolumn,subeqn,amsmath,floats,tabularx,unsortedaddress,superscriptaddress,nofootinbib,prl,amssymb,showpacs]{revtex4}
\usepackage[dvips]{graphicx}
\usepackage{epsfig}
\usepackage{tabularx}
\usepackage{color}
\usepackage[pdfpagemode=UseNone,colorlinks=true,linkcolor=blue,citecolor=blue]{hyperref}

\renewcommand\Re{\operatorname{Re}}
\renewcommand\Im{\operatorname{Im}}

\begin{document}

\title{Squeezing a thermal mechanical oscillator by stabilized parametric effect on the optical spring}

\author{A. Pontin}
\affiliation{Istituto Nazionale di Fisica Nucleare (INFN), Gruppo Collegato di Trento, I-38123 Povo, Trento, Italy}
\affiliation{Dipartimento di Fisica, Universit\`a di Trento, I-38123 Povo, Trento, Italy}
\author{M. Bonaldi}
\affiliation{Institute of Materials for Electronics and Magnetism, Nanoscience-Trento-FBK Division,
 38123 Povo, Trento, Italy}
\affiliation{INFN, Gruppo Collegato di Trento, Sezione di Padova, 38123 Povo, Trento, Italy}
\author{A. Borrielli}
\affiliation{Institute of Materials for Electronics and Magnetism, Nanoscience-Trento-FBK Division,
 38123 Povo, Trento, Italy}
\affiliation{INFN, Gruppo Collegato di Trento, Sezione di Padova, 38123 Povo, Trento, Italy}

\author{F. S. Cataliotti}
\affiliation{Dipartimento di Fisica e Astronomia, Universit\`a di Firenze, Via Sansone 1, I-50019 Sesto Fiorentino (FI), Italy}
\affiliation{European Laboratory for Non-Linear Spectroscopy (LENS), Via Carrara 1, I-50019 Sesto Fiorentino (FI), Italy}
\affiliation{INFN, Sezione di Firenze}

\author{F. Marino}
\affiliation{INFN, Sezione di Firenze}
\affiliation{CNR-INO, L.go Enrico Fermi 6, I-50125 Firenze, Italy}

\author{G. A. Prodi}
\affiliation{Istituto Nazionale di Fisica Nucleare (INFN), Gruppo Collegato di Trento, I-38123 Povo, Trento, Italy}
\affiliation{Dipartimento di Fisica, Universit\`a di Trento, I-38123 Povo, Trento, Italy}

\author{E. Serra}
\affiliation{Istituto Nazionale di Fisica Nucleare (INFN), Gruppo Collegato di Trento, I-38123 Povo, Trento, Italy}
\affiliation{Dept. of Microelectronics and Computer Engineering /ECTM/DIMES, Delft University of Technology, Feldmanweg 17, 2628 CT  Delft, The Netherlands}

\author{F. Marin}
\email[Electronic mail: ]{marin@fi.infn.it}
\affiliation{Dipartimento di Fisica e Astronomia, Universit\`a di Firenze, Via Sansone 1, I-50019 Sesto Fiorentino (FI), Italy}
\affiliation{European Laboratory for Non-Linear Spectroscopy (LENS), Via Carrara 1, I-50019 Sesto Fiorentino (FI), Italy}
\affiliation{INFN, Sezione di Firenze}

\date{\today}
\begin{abstract}
We report the confinement of an optomechanical micro-oscillator in a squeezed thermal state, obtained by parametric modulation of the optical spring. We propose and implement an experimental scheme based on parametric feedback control of the oscillator, which stabilizes the amplified quadrature while leaving the orthogonal one unaffected. This technique allows us to surpass the -3dB limit in the noise reduction, associated to parametric resonance, with a best experimental result of -7.4dB. In a moderately cooled system, our technique can be efficiently exploited to produce strong squeezing of a macroscopic mechanical oscillator below the zero-point motion.
\end{abstract}

\pacs{42.50.Wk,45.80.+r,05.40.-a,07.10.Cm}

\maketitle

A recent major breakthrough in experimental quantum mechanics is the possibility to prepare macroscopic systems close to their fundamental quantum state. In particular, micro- and nano-oscillators have been recently cooled down to an occupation number close to unity or even below it \cite{oconnell10,teufel11,chan11,verhagen12,Purdy2012}. While remaining in a thermal state, such systems display peculiar quantum properties such as asymmetric modulation sidebands induced in a probe field \cite{safavi}. A further interesting development would be the creation of a qualitatively different quantum state, for instance a mechanical squeezed state. To this purpose, possible techniques are back-action evading measurements \cite{Braginsky1980,Clerk2008,Hertzberg2009} and degenerate \cite{Rugar1991,Briant2003} or nearly-degenerate \cite{Bowen2011,Bowen2012,Bowen2013} parametric modulation.
Mechanical oscillators operate in the degenerate parametric regime when their spring constant is modulated at twice the oscillator resonance frequency.  In such regime, the response of the oscillator to  an external excitation acting close to resonance is enhanced, until the parametric modulation depth reaches a threshold marking the birth of self-oscillations (parametric resonance) \cite{landau}. More precisely, the response is amplified in the quadrature of the motion in phase with the parametric modulation, and de-amplified in the orthogonal quadrature ($\pi/2$ quadrature). Therefore, the distribution of fluctuations in the phase plane caused by stochastic excitation is squeezed, and in particular its variance is reduced below its free-running value in the $\pi/2$ quadrature. As a consequence, parametric effect can be used to produce quadrature squeezed states of a macroscopic oscillator, similarly to what is  commonly obtained for the electromagnetic field in optical parametric oscillators \cite{milburn,collet}. This effect has already been demonstrated for thermal oscillators \cite{Rugar1991,Briant2003,Bowen2013}, and is expected even for the quantum noise \cite{Bowen2011,Bowen2012}. However, since the amplified quadrature evolves into self-oscillations for an excitation strength approaching the threshold, the corresponding noise reduction in the $\pi/2$ quadrature, monotonous with the parametric excitation, is limited to -3dB. This is a general feature of parametric squeezing \cite{milburn,collet}. A proposal to surpass this limit, based on continuous weak measurements and detuned parametric drive, is reported in Ref. \cite{Bowen2011}. A recent experiment \cite{Bowen2013} shows indeed that the uncertainty in the knowledge of the oscillator trajectory in the phase space (\emph{localization}), is squeezed with a minimal variance reduced by -6.2dB with respect to that of a free thermal oscillator. The authors also suggest that, using the information on the oscillator position in an appropriate feedback loop, even the \emph{confinement} of the oscillator in a strongly ($>3$dB) squeezed state could be obtained, though such result has not yet been demonstrated.

In this work, we report on the observation of the confinement of  a micro-oscillator in a squeezed thermal state, obtained by parametric modulation of the optical spring constant \cite{sheard}. We also propose and apply an experimental scheme based on parametric feedback that, stabilizing the amplified quadrature without influencing the orthogonal one, allows to surpass the -3dB barrier on noise reduction, with a best experimental result of -7.4dB.

Our opto-mechanical system is composed of a low-deformation micro-oscillator \cite{SerraAPL2012,SerraJMM2013} with high-reflectivity coating, working as end mirror in a $0.57$~mm long, high Finesse ($\mathcal{F}=57000$, half-linewidth $\kappa/2\pi = 2.3$~MHz) Fabry-Perot cavity. The oscillator has resonance frequency $\omega_m/2\pi =128960$~Hz, mechanical quality factor $Q=\omega_m/\gamma_m=16000$ and effective mass $m = 1.35~10^{-7}$~Kg. In the presence of an input field, the intracavity power depends on the the detuning $\Delta = \omega_L-\omega_c$ between the laser and the cavity resonance, and actually on the cavity length $L_c$. As a consequence, the micro-oscillator feels a position-dependent radiation pressure that can be described as the effect of an additional \emph{optical spring} \cite{sheard}. The delay in the intracavity field buildup gives an imaginary component in the spring constant, which modifies the damping coefficient $\gamma_{\mathrm{eff}}$ of the opto-mechanical system. The complex optical spring constant is $m|G|^2 \omega_m \Delta /[(\kappa+i\omega)^2+\Delta^2]$ where $G$ is the effective opto-mechanical coupling constant with $|G|^2$ proportional to the intracavity power. For the case of our interest (\emph{bad cavity limit} $\kappa\gg \omega_m$, small detuning $\Delta\ll
  \kappa$, and $\omega\approx \omega_m$) the expression can be simplified introducing the quantities
\begin{equation}
K_{\mathrm{opt}} \approx \frac{m |G|^2 \omega_m}{\kappa^2}\,\Delta
\label{Kopt}
\end{equation}
\begin{equation}
\gamma_{\mathrm{opt}} \approx \frac{2 K_{\mathrm{opt}}}{m \kappa}
\label{gammaopt}
\end{equation}
and writing the effective susceptibility as $\,\chi_{\mathrm{eff}}^{-1}=m\left(\omega_{\mathrm{eff}}^2-\omega^2-i\omega\,\gamma_{\mathrm{eff}}\right)\,$ with $\,\,\gamma_{\mathrm{eff}}=\gamma_m+\gamma_{\mathrm{opt}}\,\,$ and
\begin{equation}
\omega_{\mathrm{eff}} = \sqrt{\omega_m^2-K_{\mathrm{opt}}/m}\simeq \omega_m - \frac{|G|^2}{2\kappa^2}\,\Delta \,.
\label{omegaeff}
\end{equation}
To our purpose, it is useful to underline that a) the frequency shift is approximately proportional to the detuning, and therefore a laser beam can be used to control it; b) the damping is also dependent on the detuning, therefore by varying the working point we can chose the effective resonance width; c) in the \emph{bad cavity} limit, the shift in the resonance frequency is larger than the variation in its width, thus the latter can be neglected when considering small variations of $\Delta$ around the working point.

\begin{figure}[ht]
 \centering
\includegraphics[width=86mm,height=121mm]{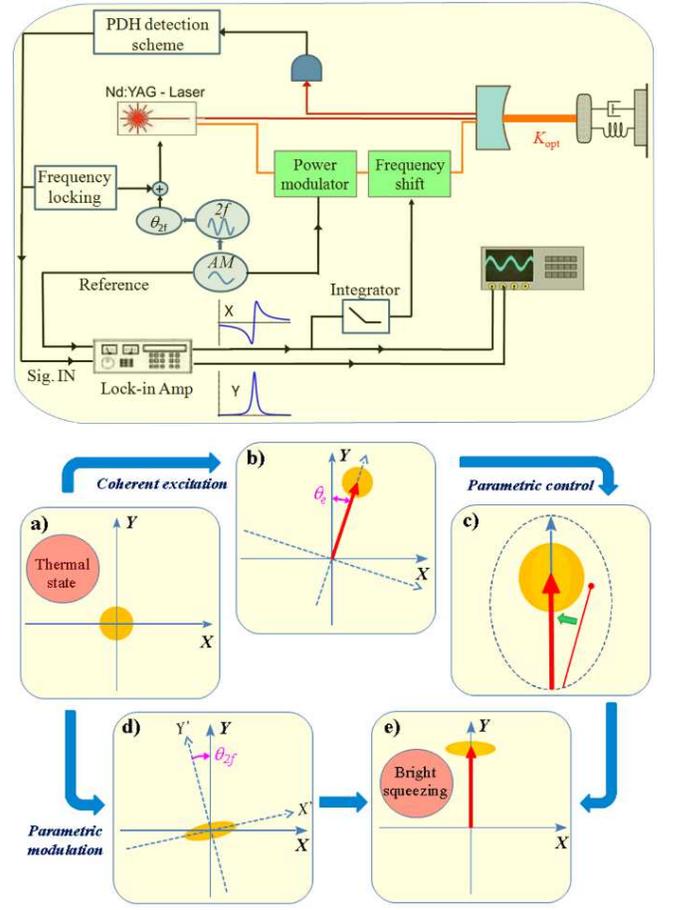}
\caption{(Color online) Upper panel:  scheme of the experimental setup. Lower panel: sketch of the experimental techniques applied to the oscillator to obtain the bright squeezed state (e) from the thermal state (a).}
 \label{setup}
\end{figure}

Our setup is sketched in Fig. \ref{setup} and more details are given in the Supplemental Material. Two laser beams derived from the same Nd:YAG source are superimposed with orthogonal polarizations and matched with an efficiency surpassing $96\%$ to a cavity mode. The first one (\emph{probe beam}, with a power of $\sim 100 \mu$W) is used in a Pound-Drever-Hall (PDH) scheme \cite{Drever} to obtain a signal proportional to the detuning. The PDH signal is used to lock the laser to the cavity resonance and to measure the oscillator displacement $x(t)$. The locking bandwidth is about 15 kHz and additional notch filters assure that the servo loop does not influence the system dynamics in the frequency region around the oscillator resonance. For the measurement, the PDH signal is sent to a double-phase digital lock-in amplifier whose outputs are acquired for the reconstruction of the motion of the oscillator. The second, more powerful beam (\emph{control beam}, input power $\sim$1mW, giving an intracavity power of $\sim 16$W and an opto-mechanical coupling at resonance $|G|^2 \simeq 6\cdot 10^{12}$Hz) has a frequency shift with respect to the probe, obtained through acousto-optic modulators, that allows to compensate the cavity birefringence and vary the detuning. This second beam is used to set and control the optical spring. An additional electro-optic power modulator in its path allows to produce a sinusoidal modulation ($AM$) in the radiation pressure, that excites the mechanical oscillator.

\begin{figure*}[ht]
\centering
\includegraphics[width=1\textwidth]{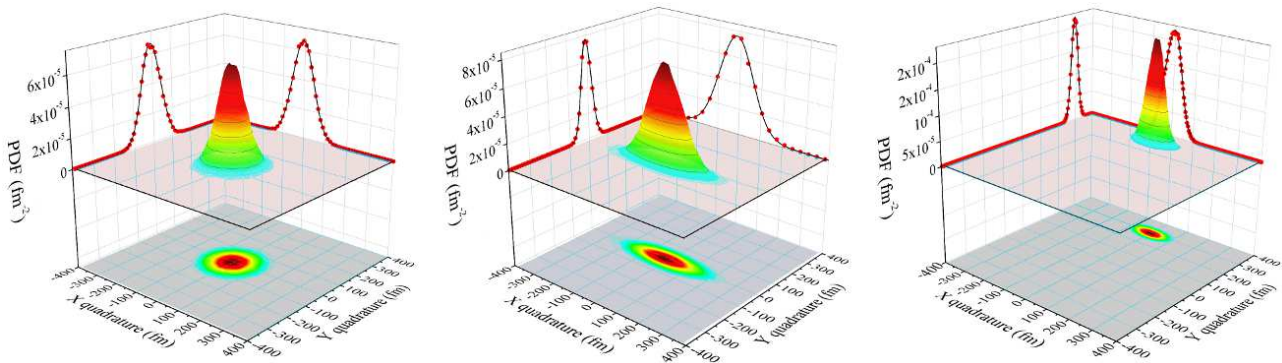}
\caption{(Color online) Phase space Probability Density Functions (PDF) for the three configurations named respectively a), d) and e) in the lower panel of Fig.\ref{setup}: from left to right, thermal oscillator (a) at the effective temperature $T_{\mathrm{eff}}\simeq$15 K ($\gamma_{\mathrm{eff}} =$110 Hz); parametrically squeezed oscillator (d), with a parametric gain $g =0.83$; squeezed oscillator with coherent excitation and frequency feedback (e), with a parametric gain $g =5.4$.}
 \label{3Ddist}
\end{figure*}

The motion of the oscillator can be decomposed into two quadratures $X(t)$ and $Y(t)$ in an arbitrary rotating frame at frequency $\omega_0$, according to $x(t) = X(t) \sin \omega_0 t + Y(t) \cos \omega_0 t$. For the opto-mechanical oscillator at temperature $T$, by choosing $\omega_0 = \omega_{\mathrm{eff}}$, $X(t)$ and $Y(t)$ are Gaussian,  stochastic, independent variables (see the sketch in Fig. \ref{setup}a) and the experimental measurements in Fig. \ref{3Ddist} left panel ) with null average, variance $\langle X^2 \rangle=\langle Y^2 \rangle=\sigma^2_0=k_B T_{\mathrm{eff}}/m\omega_{\mathrm{eff}}^2$ where the effective temperature is $T_{\mathrm{eff}} = T \,\gamma_m/\gamma_{\mathrm{eff}}$, and Lorentzian spectral densities
\begin{equation}
S_X = S_Y = \sigma^2_0 \,\frac{\gamma_{\mathrm{eff}}}{\omega^2+\gamma_{\mathrm{eff}}^2/4} \,.
\label{S0}
\end{equation}
The addition of a coherent excitation of amplitude $F_e$ at frequency $\omega_e$ (produced, in our case, by the $AM$ oscillator) shifts the distribution in the phase plane $X-Y$ by a vector rotating with angular frequency $\omega_e-\omega_0$, and in particular if $\omega_e=\omega_0$ by a constant vector with Cartesian components $\overline{X}_e=F_e \Re\left(\chi_{\mathrm{eff}}(\omega_0) \mathrm{e}^{i\theta_e}\right)$ and $\overline{Y}_e=F_e \Im\left(\chi_{\mathrm{eff}}(\omega_0) \mathrm{e}^{i\theta_e}\right)$, where $\theta_e$ is the phase between the excitation (i.e., the modulation in the intracavity power) and the detection (Fig. \ref{setup}b). Once set $\theta_e=0$ (this is experimentally performed by tuning the lock-in reference phase), $\overline{X}_e$ vs $(\omega_0-\omega_{\mathrm{eff}})$ has a dispersive shape that can be used as error signal in a parametric feedback loop. This loop acts on the detuning of the control beam in order to correct $\omega_{\mathrm{eff}}$ and keep it fixed at $\omega_0$. The distribution in the phase plane displays now a nearly-symmetric two-dimensional Gaussian shape centered at $\overline{Y}_e$ on the $y$ axis. Since the frequency $\omega_{\mathrm{eff}}$ is now locked to $\omega_0$, it can be chosen at will and, as a consequence, we can also choose the effective resonance width $\gamma_{\mathrm{eff}}$. Moreover, frequency instabilities (thermal drifts and slow fluctuations due to the effect of intracavity radiation) are corrected and, even more important, we have a precise phase reference of the coherent component of the oscillator motion.

Let us now analyze qualitatively the effect of the feedback on the two quadratures. If $\overline{Y}_e \gg \langle (Y-\overline{Y}_e)^2+X^2\rangle$, we can define a time-dependent oscillator phase $\phi=\arctan X/Y \simeq X/\overline{Y}_e$, with $\langle\phi\rangle=0$ and  $\langle\phi^2\rangle\ll 1$, and an instantaneous angular frequency $\omega_0+\dot{\phi}$. The control loop acts by correcting $\dot{\phi}$: pictorially, it rotates the fluctuating vector $(X,Y)$ forcing it to point to the $y$ direction (Fig. \ref{setup}c). As long as $\phi\ll 1$, the feedback reaction just influences the $X$ quadrature, while $Y$ remains free.

\begin{figure}[t]
 \centering
\includegraphics[width=86mm,height=107mm]{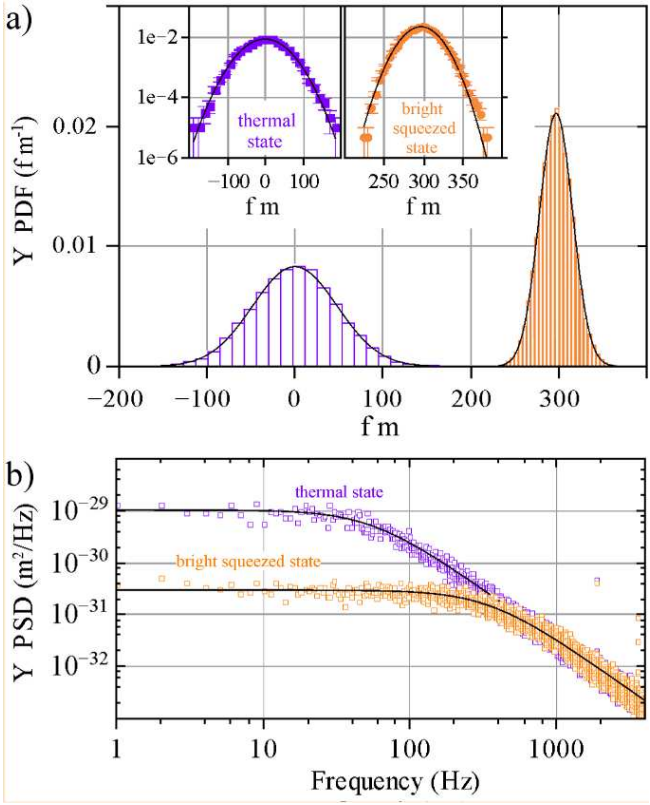}
\caption{(Color online) Upper panel: Experimental PDFs of the $Y$ quadrature. Violet (dark gray) histogram: thermal oscillator at $T_{\mathrm{eff}}\simeq$15 K.  Orange (light gray) histogram: squeezed oscillator (with coherent excitation and parametric feedback) with $g=5.9$.  Solid lines show the Gaussian fitting functions. In the inset the same histograms are shown in logarithmic scale, with statistical error bars (see Supplemental Material). Lower panel: corresponding Power Spectral Densities (PSD) of the $Y$ quadrature, with Lorentzian fitting functions.}
\label{spettri}
\end{figure}

We now move to the most important part of this work, the parametric squeezing. A modulation of the spring strength is just obtained by modulating the detuning of the control beam, according to Eq. \ref{Kopt}. The modulation signal is derived from a copy of the reference oscillator at $\omega_0$, frequency-doubled and phase-shifted by $\theta_{2f}$. For a free-running opto-mechanical oscillator (i.e., switching off the coherent excitation and frequency feedback), the expected variance in the two quadratures $X'$ and $Y'$, now referred to the phase of the parametric modulation, are \cite{Briant2003}
\begin{equation}
\sigma^2_{X'}=\langle X'^2 \rangle = \frac{\sigma_0^2}{1-g}
\label{varX'}
\end{equation}
\begin{equation}
\sigma^2_{Y'}=\langle Y'^2 \rangle = \frac{\sigma_0^2}{1+g}
\label{varY'}
\end{equation}
where $g$ is the parametric gain, proportional to the depth of the parametric modulation, and we have identified $Y'$ with the previously mentioned $\pi/2$ quadrature. The spectral densities maintain a Lorentzian shape, with width multiplied respectively by $(1-g)$ and $(1+g)$. In other words, the motion is additionally damped along the $y$ axis, and anti-damped along the $x$ axis, without modifying the input noise. In the $X-Y$ plane, we find an elliptical probability distribution rotated by an angle (Fig. \ref{setup}d) which can be set to zero by tuning $\theta_{2f}$, thus setting $X'\equiv X$ and $Y'\equiv Y$ (the experimental PDFs are shown in the central panel of Fig. \ref{3Ddist}). The variance $\sigma^2_{X'}$ clearly diverges for $g \to 1$, giving the upper limit  $\sigma^2_{Y'} < 0.5 \,\sigma_0^2$ (the mentioned -3dB limit reduction). Switching on the coherent modulation just shifts the ellipse center to $\left(0,\,\overline{Y}_e/(1+g)\right)$ (Fig. \ref{setup}e). The configuration in the phase plane is now equivalent to that of an optical field with \emph{bright squeezing} \cite{bright1}. However, we remark that the fluctuations along $x$ still increase with $g$ and the squeezing remains limited to 3dB. On the other hand, by activating the parametric feedback we can reduce the $X$ fluctuations while depressing the parametric amplification and preventing the divergence of $\langle X^2 \rangle$. As a consequence, the parametric gain $g$ can now be increased above unity (Fig. \ref{3Ddist} right panel). $S_X$ is deformed and depends on the electronic servo-loop, but the standard deviation of $X$ is maintained close to its thermal value. The crucial issue is that the $Y$ quadrature remains free. The fluctuations on $Y$ maintain indeed a Gaussian distribution and $S_Y$ keeps a Lorentzian shape, as shown in Fig. \ref{spettri}. The $Y$ variance $\sigma^2_Y=\langle\left(Y-\overline{Y}_e\right)^2\rangle$ is further reduced below the -3dB barrier, continuing to follow Eq. \ref{varY'}.

\begin{figure}[t]
 \centering
\includegraphics[width=86mm,height=58mm]{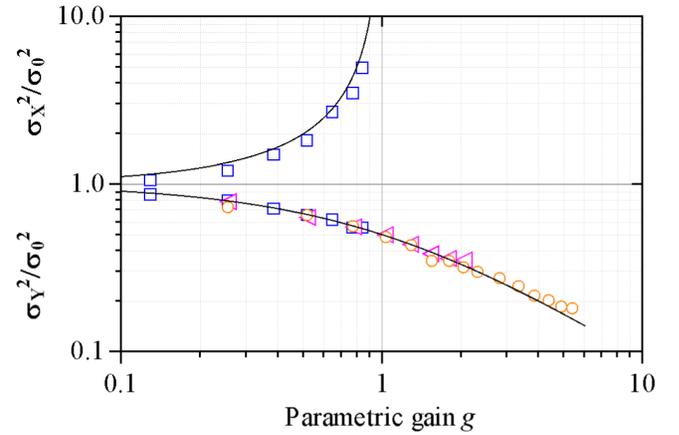}
\caption{(Color online) Normalized measured variances of the $X$ and $Y$ quadratures (see text), as a function of the parametric gain $g$. (Squares) Parametric modulation without coherent excitation and parametric feedback: $\omega_{\mathrm{eff}}/2\pi\equiv \omega_0/2\pi = 127400$~Hz ($\gamma_{\mathrm{eff}}/2\pi=160$ Hz, $T_{\mathrm{eff}} \simeq 10$ K at $g=0$). (Circles) Parametric modulation in the presence of coherent excitation and parametric feedback for $\omega_{\mathrm{eff}}/2\pi\equiv \omega_0/2\pi = 127400$~Hz ($\gamma_{\mathrm{eff}}/2\pi=160$ Hz)  and (Triangles) $\omega_{\mathrm{eff}}/2\pi\equiv \omega_0/2\pi = 128000$~Hz ($\gamma_{\mathrm{eff}}/2\pi=110$ Hz, $T_{\mathrm{eff}} \simeq 15$ K at $g=0$). Solid lines represent the theoretical curves $1/(1-g)$ and $1/(1+g)$.}
 \label{varianze}
\end{figure}

This is shown quantitatively in Fig. \ref{varianze}, where we plot the variances for the $X$ and $Y$ quadratures, normalized to their free-running value in the absence of parametric modulation, for two experimental configurations: the parametrically squeezed oscillator, and the system with coherent excitation and parametric feedback. 
In both cases, the amplitude of the coherent excitation has been adapted during the measurement in order to keep a constant value of the coherent component in the oscillator motion, i.e., a constant $\langle Y \rangle\simeq \overline{Y}_e$, compensating the parametric de-amplification. This value is $\langle Y \rangle\simeq 300$fm, i.e., at least 6 times larger than the standard deviation of the thermal distributions. This assures, together with the stabilization of the $X$ quadrature, that the condition $\phi\ll 1$ is satisfied. The solid lines in Fig. \ref{varianze} are given by the expressions $1/(1-g)$ and $1/(1+g)$ (see Eqs. \ref{varX'},\ref{varY'}), with $g = V_{2f}/V_{th}$, where $V_{2f}$ is the amplitude of the modulation sent to the laser frequency controller, and the threshold $V_{th}$ is obtained by fitting Eq. \ref{varY'} to the variance of $Y$. The maximum noise reduction is -7.4$\pm$0.2dB, limited by the appearance of instabilities in the control loop (servo bumps). An optimization of the control loop, not yet performed, would likely allow a wider working range and stronger squeezing.

In summary, we have described three original experimental results. The first one is the parametric excitation obtained by modulation of the optical spring, that produces a squeezed probability distribution in the phase plane of a thermal oscillator. This technique can be applied to a wide range of opto-mechanical systems, without additional external force. The second one is the frequency locking and phase stabilization of the opto-mechanical oscillator using parametric feedback on the optical spring constant. The addition of a known coherent component $\langle Y \rangle$ to the motion of the oscillator is crucial because it allows to establish a correspondence between quadrature fluctuations (on $X$ and $Y$) and, respectively, \emph{phase} and \emph{amplitude} fluctuations, just like in the optical case. This situation facilitates the theoretical modeling and, more importantly, it allows to limit the effect of parametric feedback, acting on the oscillator phase/frequency, only to one quadrature at first order. Third issue, we have demonstrated noise reduction and quadrature confinement below the -3dB barrier by means of a parametric feedback scheme. Again the last two results can be reproduced in a large variety of opto-mechanical devices, and would become particularly meaningful in systems with reduced effective mass, which can be operated at low occupation number \cite{oconnell10,teufel11,chan11,verhagen12,Purdy2012}. In our configuration, the parametric feedback plays the role of the low-fidelity measurement used in Ref. \cite{Bowen2013}, and it is indeed based on a measurement of the oscillator motion (performed in our setup by the PDH signal). We remark that the back-action of this measurements influences both quadratures, and therefore sets a limit to the achievable noise reduction. However, a weak measurement (with sensitivity well below the standard quantum limit) is sufficient to confine the (classical) motion of the $Y$ quadrature, giving a limited detrimental effect. This set a similarity with the schemes analyzed in Ref.\cite{Bowen2012}. However, the combination of coherent excitation and parametric feedback loop sets an \emph{a priori} limitation to the necessary low-fidelity estimation, that is remarkably along the $Y= \langle Y \rangle$ line. This avoids the problem of the optimal filter \cite{Bowen2012,Bowen2013}. Finally, we notice that, due to the strong obtainable squeezing, the starting point can be a moderately cooled oscillator (with occupation number significantly above unity), that can even be reached in the \emph{bad cavity} configuration exploited in this work. As a consequence, our scheme can be efficiently exploited to produce a macroscopic mechanical oscillator in a \emph{bright squeezed} state.

F.M. thanks M. Prevedelli for the discussion on phase locking. This work has been supported by the European Commission (ITN-Marie Curie project cQOM), by MIUR (PRIN 2010-2011) and by INFN (HUMOR project).


\begin{thebibliography}{99}
\bibitem{oconnell10} A. D. O'Connell , {\it et al.}
 Nature {\bf 464}, 697-713 (2010).
\bibitem{teufel11} J. D. Teufel, {\it et al.}
 Nature {\bf 475}, 359-363 (2011).
\bibitem{chan11} J. Chan,
{\it et al.}
 Nature {\bf 478}, 89-92 (2011).


\bibitem{verhagen12} E. Verhagen , S. Del\'eglise , S. Weis , A. Schliesser,  \& T. J. Kippenberg,
{\it Nature} {\bf 482}, 63-67 (2012).
\bibitem{Purdy2012} T. P. Purdy, R. W. Peterson, P.-L. Yu, and C. A. Regal,
New J. Phys. {\bf 14}, 115021 (2012)

\bibitem{safavi} A. H. Safavi-Naeini, J. Chan, J. T. Hill, T. P.
 Mayer Alegre, A. Krause, and O. Painter, Phys. Rev. Lett. {\bf 108},
 033602 (2012).
\bibitem{Braginsky1980}V. B. Braginsky, Y. I. Vorontsov, K. S. Thorne, Science {\bf 209}, 547 (1980).
\bibitem{Clerk2008}A. A. Clerk, F. Marquardt and K. Jacobs, New J. Phys. {\bf10}, 095010 (2008).
\bibitem{Hertzberg2009}J. B. Hertzberg, T. Rocheleau, T. Ndukum, M. Savva, A. A. Clerk and K. C. Schwab, Nature Phys. {\bf 6}, 213 (2009).
\bibitem{Rugar1991}D. Rugar and P. Gr\"utter, Phys. Rev. Lett. {\bf 67}, 699 (1991).
\bibitem{Briant2003}T. Briant, P. F. Cohadon, M. Pinard and A. Heidmann, Eur. Phys. J. D {\bf 22}, 131 (2003).
\bibitem{Bowen2011}A. Szorkovszky, A. C. Doherty, G. I. Harris, and W. P. Bowen, Phys. Rev. Lett. {\bf 107}, 213603 (2011).
\bibitem{Bowen2012}A. Szorkovszky, A. C. Doherty, G. I. Harris, and W. P. Bowen, New J. Phys. {\bf 14}, 095026 (2012).
\bibitem{Bowen2013}A. Szorkovszky, G. A. Brawley, A. C. Doherty, and W. P. Bowen, Phys. Rev. Lett. {\bf 110}, 184301 (2013).
\bibitem{landau} L. D. Landau and E. M. Lifshitz, {\it Mechanics }, chap. 5,
 page 80 (Pergamon Press, Oxford, 1976).
\bibitem{milburn}D. F. Walls, G. J. Milburn, {\it Quantum Optics} (Springer-Verlag, Berlin, 2008).

\bibitem{collet} M. J. Collet and C. W. Gardiner, Phys. Rev. A {\bf
 30}, 1386 (1984).
\bibitem{sheard} B. S. Sheard, M. B. Gray, C. M. Mow-Lowry, D. E. McClelland, and S. E. Whitcomb Phys. Rev. A {\bf 69}, 051801(R) (2004); V.B. Braginsky, M.L. Gorodetsky and F.Ya. Khalili, Phys. Lett. A  {\bf 232}, 340 (1997).
\bibitem{SerraAPL2012}E. Serra, A. Borrielli, F. S. Cataliotti, F. Marin, F. Marino, A. Pontin, G. A. Prodi and M. Bonaldi, Appl. Phys. Lett. {\bf 101}, 071101 (2012).
\bibitem{SerraJMM2013}E. Serra, A. Bagolini, A. Borrielli, M. Boscardin, F. S. Cataliotti, F. Marin, F. Marino, A. Pontin, G. A. Prodi, M. Vannoni and M. Bonaldi, J. Micromech. Microeng. {\bf 23}, 085010 (2013).
\bibitem{Drever} R.W.P.~Drever {\it et al.}, Appl. Phys. B {\bf 31}, 97 (1983).

\bibitem{bright1}F. A. M. de Oliveira, and P. L. Knightm Phys. Rev. Lett. {\bf 61}, 830 (1988); R. Paschotta, M. Collett, P. Kiirz, K. Fiedler, H. A. Bachor, and J. Mlynek, Phys. Rev. Lett. {\bf 72}, 3807 (1994).


\end{thebibliography}
\end{document}


\title{Supplemental information to: "Squeezing a thermal mechanical oscillator by stabilized parametric effect on the optical spring"}

\maketitle

\section{Experimental setup}

\begin{figure}[h]
 \centering
\includegraphics[width=0.9\textwidth]{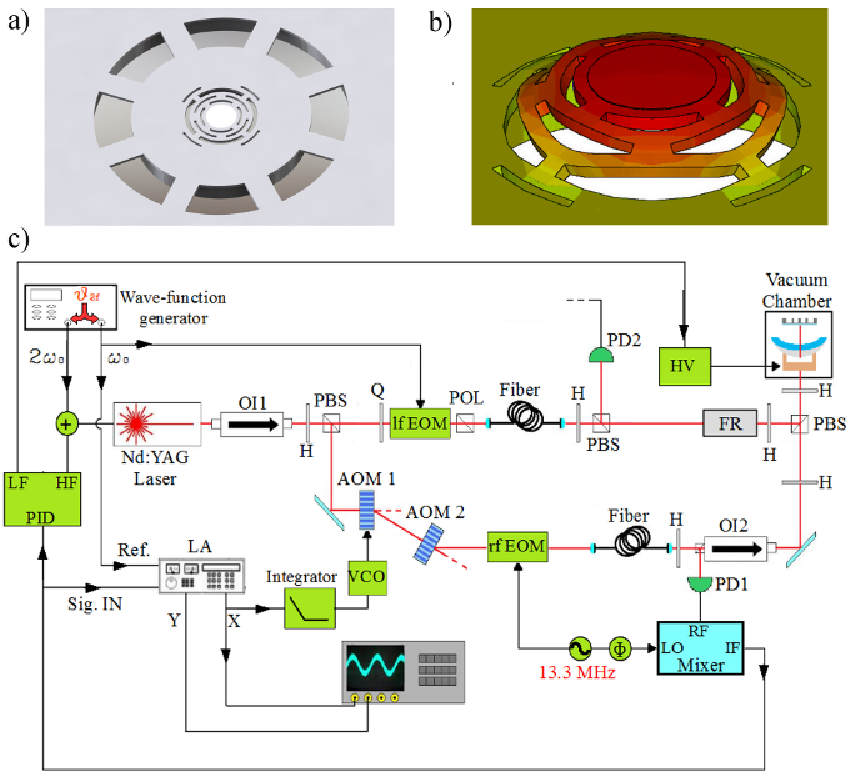}
\caption{(Color online) a) CAD rendering of the micro-oscillator, showing the intermediate vibrational filter supporting the central structure with the mirror. The diameter of the central mirror is 620$\mu$m. b) Image produced by FEM, showing the shape of the fundamental mechanical mode under study. The displacement is maximum in the red (dark gray) central mirror. c) Scheme of the experimental apparatus. Optical isolator (OI); acousto-optic modulator (AOM); electro-optic modulator (EOM); half-wave plate (H); quarter-wave plate (Q); polarizing beam splitter (PBS); polarizer (POL); Faraday rotator (FR); photodiode (PD); lock-in amplifier (LA); voltage controlled oscillator (VCO); delay line for phase control ($\phi$); high voltage amplifier (HV). Black lines indicate the electronic part of the setup. We have omitted for clearness in the scheme an additional Michelson interferometer, monitoring the displacement of the micro-oscillator from its back side, and used for mechanical characterizations of the free oscillator.}
 \label{fig:appa}
\end{figure}

\begin{figure}[h]
 \centering
\includegraphics[width=0.475\textwidth]{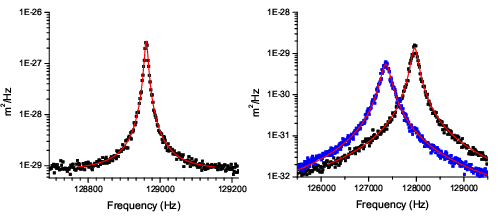}
\caption{(Color online) Left panel: spectrum of the thermal displacement of the free-running oscillator, acquired by means of a Michelson interferometer. Right panel: displacement spectra, obtained from the PDH signal, for two different values of the detuning of the control beam, showing the optical spring effect. Black squares: $\omega_{\mathrm{eff}}/2\pi = 128000$~Hz , $\gamma_{\mathrm{eff}}/2\pi=110$ Hz, $T_{\mathrm{eff}} \simeq 15$ K. Blue squares: $\omega_{\mathrm{eff}}/2\pi = 127400$~Hz, $\gamma_{\mathrm{eff}}/2\pi=160$ Hz, $T_{\mathrm{eff}} \simeq 10$ K. Solid red lines are Lorentzian fitted curves.}
 \label{spettri_osc}
\end{figure}

The mechanical oscillator used in this experiment (Fig.~\ref{fig:appa}a) has been realized in the Microfabrication Facility of FBK \cite{fbk} and belongs to a class of Micro-Opto-Mechanical-System specifically developed for having at the same time low optical and mechanical losses \cite{SerraJMM2013}. The reflective coating is obtained by the deposition of alternate Ta$_2$O$_5$/SiO$_2$ quarter-wave layers for a total thickness of about 5.9 $\mu$m. Optical cavities with Finesse up to 65\,000 have been realized with these devices, depending on the local wafer waviness \cite{SerraPRA2012}. A structure made of alternate torsional and flexural springs supports the central mirror and allows its vertical displacement with a minimal internal deformation (see Fig.~\ref{fig:appa}b). This feature reduces the mechanical loss in the optical coatings and allows to reach quality factors up to 10$^5$ at cryogenic temperatures. In this specific case the measurements were performed at room temperature, where the thermoelastic loss in the flexural springs prevents from achieving quality factors much surpassing $\sim10^4$ \cite{SerraAPL2012}. The particular structure used in the present work has resonance frequency $\omega_m/2\pi =128960$~Hz, mechanical quality factor $Q=\omega_m/\gamma_m=16000$ and effective mass $m = 1.35~10^{-7}$~Kg. The mechanical parameters are measured from a spectrum of the thermal motion acquired by means of a Michelson interferometer, monitoring the position of the oscillator from its uncoated back-side (Fig. \ref{spettri_osc}a). The effective mass is deduced by assuming thermal equilibrium at room temperature, and it is in agreement with the value estimated from FEM simulations, and with the value deduced from the response to a calibrated excitation applied by a modulated radiation pressure \cite{Decumis}.

The micro-oscillator is used as end mirror of a $0.57$~mm long Fabry-Perot cavity formed with a $50$~mm radius silica input mirror (transmissivity $\sim 50$~ppm), operating in a vacuum chamber at $10^{-3}$~Pa. The input coupler is glued on a piezoelectric transducer (PZT) for coarse cavity tuning. The cavity Finesse is $\mathcal{F}$~=~57\,000, (half-linewidth $\kappa/2\pi = 2.3$~MHz), and operates slightly under-coupled (the ratio between input mirror transmission and total roundtrip losses is $\sim 0.45$). A suspension system in the vacuum chamber isolates the cavity from environmental vibrations.

The experimental setup is sketched in Fig.~\ref{fig:appa}c. The light source is a cw Nd:YAG laser operating at $\lambda$=1064~nm. After a 40~dB optical isolator (OI1), the laser radiation is split into two beams. The first one (\emph{probe beam}, power $\sim 100\mu$W measured at the cavity input) is frequency shifted by means of two acousto-optic modulators (AOM) operating on opposite diffraction orders. A resonant electro-optic modulator (rf EOM) provides phase modulation at 13.3~MHz used for the Pound-Drever-Hall (PDH) detection scheme \cite{Drever}. The locking bandwidth is about 15 kHz and additional notch filters assure that the servo loop do not influence the system dynamics in the frequency region around the oscillator frequency. The servo loop acts on the cavity PZT to correct slow drifts (above $\sim1$s) and on a PZT installed on the laser crystal for faster corrections. The second, more powerful beam (\emph{control beam}, power $\sim$1mW at the cavity input), is employed to induce a strong optomechanical coupling and control the optical spring. A power modulator formed by a quarter-wave plate, an electro-optic modulator (lf EOM) and a polarizer allows to produce a sinusoidal modulation in the intensity of the control beam, and actually on the radiation pressure exciting the oscillator. The maximum modulation depth used in the experiment is 0.5$\%$ peak-to-peak, giving an intracavity modulated force with amplitude $F_e^{\mathrm{max}}\simeq 2 \cdot 10^{-10}$N.

Both beams are sent to the second part of the apparatus by means of single-mode, polarization maintaining optical fibers, overlapped with orthogonal polarizations in a polarizing beam-splitter and then mode-matched to the optical cavity with an efficiency above 96$\%$. Since the cavity is birefringent (around 28 MHz), an half-wave plate is used to align the linear polarizations with the cavity polarization axes. The AOM shifters allow to compensate for the difference between the optical resonance frequencies for the two polarizations and set the detuning of the control beam. The reflected probe beam, on its back path, is deviated by the input polarizer of a second optical isolator (OI2) and collected by a photodiode (PD1) for the PDH detection and laser frequency locking. The reflected control beam, whose polarization is rotated by a double pass through a Faraday rotator (FR), is monitored by the photodiode~PD2.

The PDH signal is also monitored by a first digital scope and sent to a double-phase digital lock-in amplifier, whose outputs are integrated with a time constant of 80$\mu$s and acquired by a second digital scope with a resolution of 12 bit and a sampling interval of 21$\mu$s. The PDH signal, approximately linear with the probe detuning $\Delta_p$ if $\Delta_p/\kappa \ll 1$, is calibrated by means of a sinusoidal modulation at $\sim20$kHz sent to the laser PZT. The real amplitude of this modulation is directly measured during the acquisitions (since it is influenced by the frequency servo loop), while the laser response (in Hz/V) had been previously calibrated with a Michelson interferometer. The ratio between the laser frequency and the cavity length allows to convert the detuning into cavity displacement. The overall calibration has an accuracy of $\sim20\%$. The exact gain and transfer function of the lock-in are measured using white input noise, and used to correct the output spectra (due to the short time constant, the spectral correction is just relevant for frequencies above several oscillator widths). Examples of displacement noise spectra acquired by the first scope (before the lock-in demodulation), for different values of the control detuning, are shown in Fig. \ref{spettri_osc}b. The shift of the mechanical resonance is in good agreement with the expected optical spring effect (see Eq. (3) in the main text). The peak of the oscillator lies on a background displacement noise of $\sim 10^{-33}\,$m$^2$/Hz, due to the other mechanical modes involved in the cavity system, while the detection noise (mostly due to electronics and probe shot noise) is at least 20dB lower. Noise spectra after the lock-in, corrected for its transfer function, are shown in Fig. 3 of the main article.

For the control of the opto-mechanical resonance, the X output of the lock-in is integrated and sent to the VCO driver of AOM1. The control bandwidth is well below 1 kHz, much lower than the locking bandwidth of the laser to the cavity, therefore we can assume that, even if the AOM shifts the probe frequency, the probe beam remains resonant with the cavity while the detuning of the control beam is modified by the integrated X signal. The effect is a stabilization of the opto-mechanical resonance, forced to coincide with the lock-in reference, by means of the optical spring tuned by the control beam.

For achieving a degenerate parametric regime, a frequency doubled copy of the reference oscillator, with controllable amplitude and phase, is sent to the laser PZT. As a consequence, both beams are frequency modulated, giving an overall modulation of the optical spring (mainly due to the more powerful control beam).

\section{Data analysis}

The spectra of the X and Y time series, containing typically $\sim10^6$ data points, are calculated using fast Fourier transform. Two examples of such spectra for the Y quadrature, corrected for the low-pass filter of the lock-in, are shown in Fig. 3b of the main text. When the parametric control of the oscillator frequency is activated, the Y spectra display a Lorentzian behavior, as shown by the fitting curve in the same Fig. 3b. The spectra of the time series taken without frequency control are better fitted by Voigt profiles, due to the slow fluctuations of the opto-mechanical frequency. The fits allow to deduce the correlation times of each time series.

For calculating the histograms shown in Fig. 3a of the main text we have first applied to the time series an additional digital low-pass filter at $\sim1$kHz to reduce the effect of spurious peaks that appear above 5kHz. This cutoff is still well above the half-width of the opto-mechanical resonance, that even with the largest parametric gain is around 300 Hz. We have then reduced the number of samples by keeping one data point every correlation time, in order to obtain sets of uncorrelated samples that have been used to calculate the probability distributions (PDFs). Error bars reflect the standard error on each bin. The reduced $\chi^2$ of a Gaussian fit is around unity for both the PDFs of the thermal state and that of the Y quadrature in the squeezed state, indicating that even in the latter case the fluctuations are stochastic with Gaussian-Lorentzian statistics.

The variance of the Y output (as well as the one of the X output when relevant) is calculated from the integral of the spectra. The statistical uncertainty calculated from several independent time series is typically around $3\%$. The absolute calibration of the oscillator displacement has an accuracy of just $\sim20\%$, but for the squeezing factor (the ratio between the Y variance in the squeezed  and thermal states) the estimated uncertainty is 0.2dB.